\documentclass[fleqn,usenatbib]{mnras}

\usepackage{newtxtext,newtxmath}

\usepackage[T1]{fontenc}
\usepackage{ae,aecompl}

\usepackage{graphicx}	
\usepackage{amsmath}	
\usepackage{amssymb}	
\usepackage{color}

\newcommand{\msol}{M$_{\odot}$}

\newcommand{\Gaia}{\textsl{Gaia}}

\title[Weighing in on black hole binaries with BPASS]{Weighing in on black hole binaries with BPASS: LB-1 does not contain a 70M$_{\odot}$ black hole}

\author[J.J. Eldridge et al.]{J.J. Eldridge,$^{1}$\thanks{E-mail: j.eldridge@auckland.ac.nz}
E.R. Stanway,$^{2}$
K. Breivik,$^{3}$
A.R. Casey,$^{4,5}$
D.T.H. Steeghs,$^{2}$ and \newauthor
H. F. Stevance$^{1}$ 
\\
$^{1}$Department of Physics, University of Auckland, Private Bag 92019, Auckland, New Zealand\\
$^{2}$Department of Physics, University of Warwick, Gibbet Hill Road, Coventry, CV4 7AL, UK\\
$^{3}$Canadian Institute for Theoretical Astrophysics, University of Toronto, 60 St. George Street, Toronto, Ontario, M5S 1A7, Canada\\
$^{4}$Center of Excellence for Astrophysics in Three Dimensions (ASTRO-3D), Australia\\
$^{5}$School of Physics and Astronomy, Monash University, Victoria 3800, Australia\\
}

\date{Accepted XXX. Received YYY; in original form ZZZ}

\pubyear{2015}
 \hypersetup{draft} 
\begin{document}
\label{firstpage}
\pagerange{\pageref{firstpage}--\pageref{lastpage}}
\maketitle

\begin{abstract}
The recent identification of a candidate very massive (70\,\msol) black hole is at odds with our current understanding of stellar winds and pair-instability supernovae.
We investigate alternate explanations for this system by searching the BPASS v2.2 stellar and population synthesis models for those that match the observed properties of the system. 
We find binary evolution models that match the LB-1 system, at the reported \Gaia\ distance, with more moderate black hole masses of 4 to 7~M$_{\odot}$. We also examine the suggestion that the binary motion may have led to an incorrect distance determination by \Gaia. We find that the \Gaia\ distance is accurate and that the binary system is consistent with the observation at this distance. Consequently it is highly improbable that the black hole in this system has the extreme mass originally suggested. Instead, it is more likely to be representative of the typical black hole binary population expected in our Galaxy.
\end{abstract}




\section{Introduction}

The recent identification by \citet{LB1} of a candidate very massive black hole (BH) of 70\,M$_{\odot}$
{with a solar metallicity stellar companion} is a remarkable discovery. Our knowledge of stellar-mass BHs to date has come from interacting binaries, gravitational microlensing and gravitational wave events \citep[e.g.][]{2010ApJ...725.1918O,2016MNRAS.458.3012W,2019PhRvX...9c1040A}. However, this new BH 
{candidate} binary, that of \citet{2019Sci...366..637T} and those found by \citet{2018MNRAS.475L..15G,2019A&A...632A...3G} are significant discoveries and represent a new window into a population of non-interacting binaries with BH primaries through the observation of the motion of a still-living stellar companion.

The suggestion that a 70\,M$_{\odot}$ BH exists relatively nearby, within our Galaxy, is at extreme odds with the expected nature of such systems, which are predicted to occur only at very low metallicities.
Consequently, based on current stellar evolution models and understanding of stellar wind mass-loss rates and pair-instability supernovae, such a large mass for the LB-1 black hole mass is impossible, as discussed by \citet{2020ApJ...890..113B}. 
In fact, at Solar metallicity, stellar winds are strong enough to reduce all stellar progenitors of BHs to masses, at birth, to 10 to 20\,M$_{\odot}$ \citep[e.g][]{2004MNRAS.353...87E,2010ApJ...725.1918O,2010ApJ...715L.138B,2010MNRAS.408..234M,2011ApJ...741..103F,2016MNRAS.462.3302E,2017PASA...34...58E}.

\citet{LB1} propose three primary arguments in favour of the massive BH interpretation. First, that broad H$\alpha$ line emission arises from a Keplerian disk close to the hidden compact object, giving a velocity half-amplitude of $K_A=6.4\pm0.4$\,km\,s$^{-1}$ viewed at low inclination of $\sim$15$^\circ$ to the line of sight. Second, that the most likely companion is a 8.5\,M$_{\odot}$ main-sequence star, which is identified as having a velocity half-amplitude of $K_B=52.8\pm0.7$\,km\,s$^{-1}$. Third, that the source parallax reported by \Gaia\ \citep{Brown:2018} is incorrect by a factor of two, placing the system at $4.23 \pm 0.24$\,kpc instead of $2.14^{+0.51}_{-0.35}$\,kpc.

The first of these points is dependent on the correct interpretation of the H$\alpha$ line profile, a subject which the authors recognise as problematic, due to the complications that can arise from hotspots, asymmetric structures and non-disk flows lying between the binary components. This is particularly true since there is no clear evidence that the compact object is actively accreting and so the presence of a relaxed, symmetric, accretion disk that reliably traces the BH is not obvious. Even in well-established accreting BH systems during quiescent states, the H$\alpha$ distribution is usually complex \citep[e.g.,][]{2010A&A...516A..58G} and does not generally reliably track the BH \citep[e.g.,][]{2008MNRAS.384..849N}.
\citet{LB1} argue that some of these complications can be avoided by sampling the line wings, where the phasing appears to be correct. However, it is not clear whether the phase was a free parameter in the presented fits, how consistent this phase is, nor how well it is constrained.  Here again we note that in cases where $K_{A}$ can be determined through other means, such semi-amplitudes are often not reliable measures of $K_{A}$ \citep[e.g][]{1994MNRAS.266..137M,2015MNRAS.452L..31L}. Further more \citet{2020MNRAS.493L..22E} and \citet{2019arXiv191204092A} have also discussed problems with this interpretation.

The second point is the result of stellar evolution modelling. \citet{LB1} do consider the possibility of a lower mass BH, but argue that the thermal timescale of the companion star is too short for us to have observed a star in such an evolutionary state. They state that ``\textit{No natural stellar models would be consistent with such a companion..."}. This statement, however, is not substantiated by a search through the output models of binary evolution codes in \citet{LB1}. 

{The third point argues that the distance reported by \Gaia\ \citep{Lindegren:2018,Brown:2018} is incorrect by a factor of two due to a biased parallax measurement from the astrometric wobble induced by the BH, thus making the distance seem closer.} This would require \Gaia\ to have observed LB-1 at particular times of its orbit. To examine if this is the case requires non-trivial calculations.

In this paper, we investigate plausible 
{explanations} for this system by searching through the extensive BPASS\,v2.2~\footnote{Binary Population and Spectral Synthesis; \url{https://bpass.auckland.az.nz} and \url{https://warwick.ac.nz/bpass} } grid of detailed binary stellar evolution models for systems matching the observed properties of both the early-type B-star 
and its unseen BH companion. 
We also explore whether the reported parallax could be 
sufficiently biased 
{to impact \Gaia's distance constraint}.

We note that several other authors have also questioned the conclusions of \citet{LB1} on the binary system LB-1 since the result was originally published and this article was first submitted. These studies have included re-examining the spectrum of the companion \citep{2020A&A...633L...5I,2020A&A...634L...7S}, the interpretation of the broad H$\alpha$ line \citep{2020MNRAS.493L..22E,2019arXiv191204092A} and the lack of plausible evolution pathways by either isolated binary evolution of that in clusters \citep{2020ApJ...890..113B,2019arXiv191204509T,2019arXiv191206022B}. Those papers that present a refined analysis of the companion star do not lead to a change of our findings below, however the abundance information provided by detailed analysis of the spectroscopy does give an extra test of our models that we discuss below.

The outline of this article is as follows: We first describe our fitting mechanism and present the results of our model search in Section \ref{sec:meth}. 
{We investigate the impact of the astrometric wobble due to binary motion on \Gaia's parallax measurement in Section~\ref{sec:distance}.} We then discuss in Section \ref{sec:disc} the expected binary population of BH binaries from BPASS and suggest that future \Gaia\ discoveries of these systems will allow us to rigidly constrain the formation of BH at the point of core collapse within stellar interiors. 

\section{Method and Results}
\label{sec:meth}

\begin{table}
	\centering
	\caption{Summary of the best fitting values using the strongly constrained values. The values from our weaker constraints are identical within the uncertainties.}
	\label{tab:results}
	\begin{tabular}{cccc} 
	    \hline
	    \multicolumn{2}{c}{Initial Parameters}&  \multicolumn{2}{c}{Current Parameters} \\
		\hline
\multicolumn{4}{c}{Averages from models matching with 1$\sigma$.}\\
        \hline
        $M_{\rm i,com}/M_{\odot}$  &  5.39  $\pm$   0.49 & $M_{\rm com}/M_{\odot}$  & 1.03 $\pm$ 0.14\\
        $M_{\rm i,BH}/M_{\odot}$  &  3.87   $\pm$  0.90  & $M_{\rm BH}/M_{\odot}$&  8.22 $\pm$   1.24 \\
        $\log(P_i/{\rm days})$ &   0.68   $\pm$  0.10 &   $\log(P_{\rm fit}/{\rm days})$ &  1.87  $\pm$   0.07\\
        && $\log(L/L_{\odot})$ &  3.51 $\pm$   0.15 \\
        && $\log({\rm age/ yrs})$ &   7.97  $\pm$   0.10 \\
\hline
\multicolumn{4}{c}{Averages from models matching with 3$\sigma$.}\\
\hline
        $M_{\rm i,com}/M_{\odot}$  &  4.27  $\pm$   0.62 & $M_{\rm com}/M_{\odot}$  & 0.77 $\pm$ 0.14\\
        $M_{\rm i,BH}/M_{\odot}$  &  4.60   $\pm$  0.44  & $M_{\rm BH}/M_{\odot}$&  5.37 $\pm$   1.58 \\
        $\log(P_i/{\rm days})$ &   1.54   $\pm$  0.42&    $\log(P_{\rm fit}/{\rm days})$ &  1.87  $\pm$  0.07\\
        && $\log(L/L_{\odot})$ &  3.20 $\pm$   0.24 \\
        && $\log({\rm age/ yrs})$ &   8.23  $\pm$   0.13 \\
   		\hline
	\end{tabular}
\end{table}



\begin{figure*}
	\includegraphics[width=2\columnwidth]{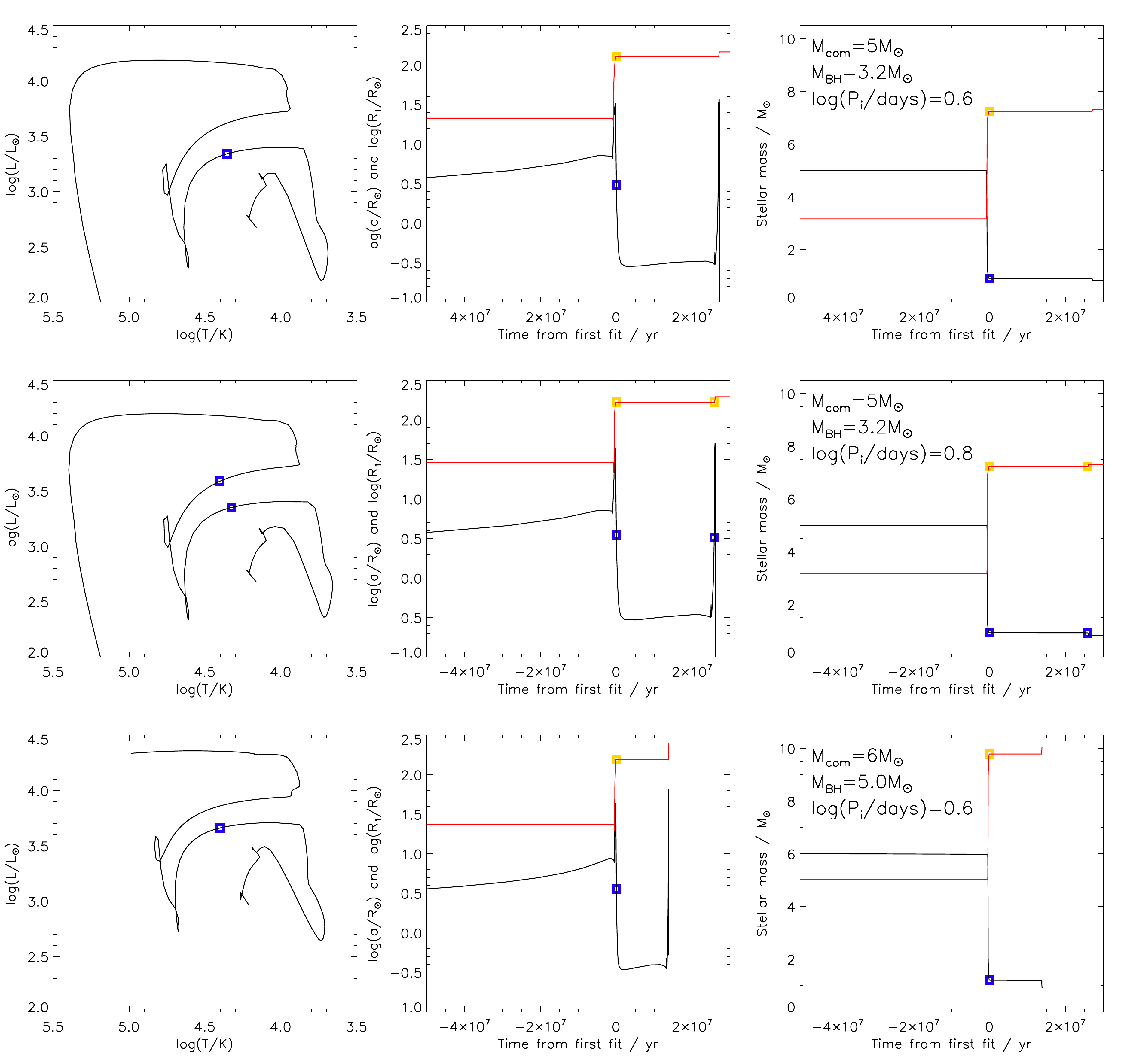}
    \caption{Panels showing the evolution of binary models that give rise to the observed LB-1 system. In the panels the lines represent the evolutionary track from the models while the squares represent the values when the system matches. Left panels are the Hertzsprung-Russell diagram for the three best fitting models, blue points are points of agreement with observed system. Central panels show the radius evolution of the companion, $R_1$, (black/blue) and orbit, $a$, (red/orange) and the right panels show the mass of the BH (red/orange) and companion (black/blue).}
    \label{fig:evolution}
\end{figure*}

\subsection{Numerical Method}

We search the BPASS v2.2 grid of binary evolution models \citep[for full details see][]{2017PASA...34...58E,2018MNRAS.479...75S} at a metallicity of $Z=0.020$, which is appropriate for Solar metallicity, as derived from spectroscopy of the B-star companion in LB-1 \citep{LB1}. We use our fiducial BPASS initial parameter distribution (see \citealt{2018MNRAS.479...75S}) adopting the broken power-law  initial mass function denoted ``imf135-300'', which has a low mass slope from \citet{1993MNRAS.262..545K} and a \citet{1955ApJ...121..161S} slope for massive stars, and extends to a zero age main sequence upper mass limit of 300\,M$_\odot$.
{The initial distributions for binary fraction, orbital period and mass ratio are drawn from the analysis of \citet{2017ApJS..230...15M}.}

We consider all binary stellar models that have already formed a BH, and record those that match the following parameters for the companion star  \citep[as derived by][]{LB1}:
\begin{enumerate}
    \item $\log (g/{\rm cm\, s^{-2}}) = 3.43\pm0.15$,
    \item $\log(T_{\rm eff}/K) = 4.258 \pm 0.15$,
    \item $M_{\rm BH}/M_{\rm com} = 8.25 \pm 1.04$,
    \item $(P/{\rm days}) = 78.9 \pm 20.0$,
    \item $\log(L/L_{\odot}) \ge 3.0$,
    \item $X_{\rm surf} \ge 0.1$,
    \item $M_{\rm BH} \ge 2.5M_{\odot}$.
\end{enumerate}

We note that BPASS models start with a range of initial orbital periods separated by 0.2\,dex intervals \citep[with the model weight based on the distribution of][]{2017ApJS..230...15M}. As a result, the final orbital period parameter space is not fully sampled, and some periods fall into gaps between adjacent models. To accommodate this, we permit a larger range for the orbital period than that determined from observations. We note that \citet{2020A&A...633L...5I} and \citet{2020A&A...634L...7S} both re-analysed the stellar spectrum and found slightly different values for the surface gravity and effective temperature, however they agree within the uncertainties and our results do not vary when we use their values. 

\subsection{The matching stellar model parameters}

Three stellar models are found to lie within the $1\,\sigma$ uncertainty range of all observed constraints on the binary parameters\footnote{We note that at Solar metallicity the BPASS population is constructed from a total of 18118 stellar models, 5168 of which are of binary stars systems including a compact remnant.}. These have initial parameters ($M_{\rm com}/M_{\odot}$, $M_{\rm BH}/M_{\odot}$, $\log(P_{\rm i}/{\rm days})$) = (5, 3.2, 0.6), (5, 3.2, 0.8) and (6, 5.0, 0.6). A further five models match all observationally-derived constraints to within their $3\,\sigma$ uncertainty range. 
{These models all have a substantially lower mass for the B-star companion than the mass reported by \citet{LB1}, but are still consistent with the observed surface gravity, due to an interaction between the B-star and BH which also circularizes the orbit.} We note that all our models assume circular orbits throughout their evolution. 
\citet{LB1} ruled out such a system due to not finding an evolutionary track that could produce it, as well as assuming the evolutionary timescale was based on an estimated thermal timescale.

We also consider a looser subset of constraints,  requiring only points (i), (ii), (iv), (v) and (vii) above: this removes constraints -- namely the binary mass ratio, surface hydrogen abundance and assumed distance -- derived from model-dependant interpretation of the H$\alpha$ line dynamics, the companion spectra and line of sight extinction.
We find 42 (203) model that match these relaxed constraints within their 1\,$\sigma$ ($3\,\sigma$) uncertainty ranges. These have similar parameters to the three systems identified above, but extend over a broader range 
{which suggests that LB-1 is likely a subset of a larger class of BH binaries}.

To constrain each property of the components of LB-1 we must account not only for the range of models which fit the data and the quality of the fit, but also for how long they remain in that state. The mean for each parameter (for instance initial mass $M_i$ or initial binary period $P_i$) is calculated by,
\begin{equation}
\langle{}A\rangle{} =\frac{ \sum_{i,j} \Delta t\,  \gamma\, A_{i,j} \prod_{k=1}^4 {p_k}}{ \sum_{i,j} \Delta t\,  \gamma \, \prod_{k=1}^4 {p_k}}
\end{equation}
where $\langle{}A\rangle{}$ is the mean of the parameter we are determining, $i$ is an index over matching stellar models, $j$ is the timestep of the model, $A_{i,j}$ is the $i$'th models $j$'th timestep value of the parameter, $\Delta t$ is the corresponding model timestep length in years, and $\gamma$ is the model weighting from the initial population synthesis parameter distribution. The term $p_k$ reflects how closely the model parameters $\alpha_{\rm k,mod}$ fit the observed constraints $\alpha_{\rm k,obs}$ and is calculated by,
\begin{equation}
p_k = \exp \left(-\frac{((\alpha_{\rm k,obs})-(\alpha_{\rm k,mod}))^2}{2 \sigma_{\alpha_{\rm k,obs}}^2}\right).
\end{equation}
We use four observational constraints  in this calculation, specifically constraints (i), (ii), (iii) and (iv).

The best fitting initial parameters of the components of LB-1 and the uncertainty in their values, as derived from matching binary stellar evolution models, are listed in Tables \ref{tab:results}. We include weighted means as described above based on all models fitting the observational parameters within 3\,$\sigma$. {We show results based on the fully constrained model sample (N=8) while noting that those from the less constrained sample (N=203) are consistent given the uncertainty on each parameter.} 
When the weaker constraints are adopted, the masses of the objects decrease and the age of the system increases.

In Figure \ref{fig:evolution} we consider the stellar evolution tracks of the 3 best-matching LB-1 models through the Hertzsprung-Russell diagram, and also the evolution of the mass of the system and the stellar and orbital radii for the best-matching model.  
At all times the masses are less than those suggested by \citet{LB1}. 
The best-matching system has an initially $\approx$ 3\,\msol\ BH with a
 5\,\msol\ companion, and evolves through Roche-lobe overflow into the binary observed today, with an approximately 1\,M$_{\odot}$ companion and a 7\,M$_{\odot}$ BH at the current time.  The other matching models show similar behaviour.
 
The evolution of the stars start with the companion on the main sequence. It interacts with the black hole shortly after the end of the main-sequence in the Hertzsprung gap. As the companion is initially more massive than the black hole, mass transfer initially shrinks the orbit until the mass ratio reverses and the orbit widens. The systems do not enter common envelope evolution. This binary evolutionary pathway has been known for some time \citep[e.g][]{1985ApJS...58..661I}. 

We note here that in our model for mass transfer onto black holes we do allow super-Eddington mass-transfer, while the Eddington Limit is assumed during accretion onto neutron stars and white dwarfs. This is not commonly assumed in other binary population synthesis calculations but we note that our models do agree with current observations of remnant masses \citep[e.g][]{2016MNRAS.462.3302E,2017PASA...34...58E}. If we were to impose the requirement that accretion onto the black hole could not exceed the Eddington rate then we expect that the initial mass of the black holes would need to increase. A good discussion and exploration of whether the Eddington limit should be applied to black hole accretion or not can be found in \citet{2020arXiv200405187V}.

For all our models shown in Figure \ref{fig:evolution}, the first time they match the LB-1 system is after the mass transfer has finished and the stellar companions are collapsing to be a sub-dwarf on the helium burning main-sequence. After completing core helium burning they re-ascend the giant branch, interacting with the black hole a second time before then continuing on to the white dwarf cooling track. Because our models match LB-1 after binary interactions we can expect the orbit to be circular, thus naturally reproducing the observed very low eccentricity for LB-1. Regardless of which scenario corresponds to the observed system, we can conclude that {the B-star} is evolving on a thermal timescale.

Finally, it is important to also understand the initial zero-age main-sequence parameters of the binary system that gave rise to our favoured models for LB-1. Here we are describing the initial parameters for when both stars in the binary are on the zero-age main-sequence, before the first supernova occurred to form the black hole. The parameters for the systems which can generates this LB-1 binary are an initial primary mass of 50 to 60\,$M_{\odot}$, a  mass ratio  $M_2/M_1 = 0.1$ (i.e. companion masses of 5 to 6~$M_{\odot}$) and an initial period $\log(P_i/{\rm days})$ in the range of 1.2 to 2. Such binaries are quite typical given the current best constraints on initial binary parameters \citep{2017ApJS..230...15M}. {The primary star typically has an evolutionary time of 4 to 4.5~Myrs and evolves to a Wolf-Rayet star via a short common-envelope evolution phase that is entered via the Darwin instability and occurs at the end of the main sequence or shortly afterwards in the Hertzsprung gap.} The star has a final mass of the order of 10~M$_{\odot}$, with this mass being almost all of the CO core and thus forming a black hole during core-collapse.

\subsection{The expected  number of LB-1 like systems}

The {total intervals during which}
the models could be observed within 1$\sigma$ of their inferred
locations in the Hertzsprung-Russell diagram are of the order of $2\times10^4$ to $8\times10^4$ years 
($1\times10^5$ to $4\times10^5$~years within 3$\sigma$)
depending on the stellar evolution model considered. Some systems,
including the one shown in Figure \ref{fig:evolution}, agree with the observed system at multiple evolutionary phases. { From the matching models we can estimate the number of systems we expect to observe that match the  constraints on LB-1. In our standard BPASS population we find that there are $0.11$ per $10^6$\,M$_{\odot}$ stars formed in the Galaxy that match all the observational constraints within their 1\,$\sigma$ uncertainty ranges. There are $1.19$ such stars per $10^6$\,M$_{\odot}$  if we relax the requirement to fit within 3\,$\sigma$ of the observed values.} When our weaker constraints are considered, we find substantially more matches to the observed parameters: 42 (203) models within 1\,$\sigma$ (3\,$\sigma$) where we expect 2.61 (13.08) per $10^6$\,M$_{\odot}$ stars formed.

Assuming a state lifetime of $8\times10^4$ years per system, and that the Milky Way has been forming stars at a steady rate of 3.5\,M$_{\odot}\,{\rm yr}^{-1}$ for the last $10^{10}$~years, these population synthesis expectations correspond to extant current-day populations of 0.03 (0.3) or 0.73 (3.7) binary systems matching these observational constraints in the Milky Way (for tightly constrained and less constrained 1\,$\sigma$ (3\,$\sigma$) matches respectively). While this is clearly a relatively rare system, it is not unreasonable to find several such objects in the Milky Way given our current understanding of stellar evolution.

\subsection{The companion luminosity, distance and extinction}

Comparing the predicted photometry from these models to that observed for LB-1 \citep{2003AJ....125.2531R,2003yCat.2246....0C} we infer distances that range from 1.60 
to 2.17\,kpc, in good agreement with the distance of $2.14^{+0.51}_{-0.35}$\,kpc derived from \Gaia\ astrometry \citep{Brown:2018,Lindegren:2018}. 
We also fit the radial velocity data reported in \citet{LB1} using \texttt{The Joker}\footnote{In the course of this work we learned that Adrian Price-Whelan (Flatiron Institute) also performed this analysis independently: https://gist.github.com/adrn/c4fd0515f95f45926c1bad6c9c09089c} \citep{2017ApJ...837...20P,astropy:2013, astropy:2018} to determine the minimum mass of the BH based on fits to the binary mass function and an assumed companion mass of $0.77\,\rm{M_{\odot}}$. We find a minimum BH mass of $M_{\rm{BH,min}} = 2.1\pm{0.93}\,M_{\odot}$. If we assume that the BH mass is $5.37\,\rm{M_{\odot}}$, as in our best fit, we derive an inclination of $21^{\circ}$, which agrees with the low inclinations derived in \citet{LB1} to within $\sim15-33\%$.

The extragalactic dust extinction along the line of sight towards LB-1 is 
$A_V=2.1$ or E(B-V)$ = 0.66$ \citep{2011AAS...21743442S}. The extinction required
to get an SED that is consistent with the spectrum inferred effective temperature and gravity is $M_V$ of 1.90 or an $E(B-V)$ of 0.61, which is 
similar to that found by \citet{LB1} of $E(B-V)$=0.55$\pm$0.03. \citet{LB1} found this extinction was consistent with the value derived from the Pan-STARRS 3D Galactic dust map \citep{2018MNRAS.478..651G} at the greater distance. 
{At the location of LB-1, this map suggests that reddening values between E(B-V)=0.22 and E(B-V)=0.40 are expected at the distances best-fitting our models. This suggests there may also be local extinction associated with the stellar source. We note that fixing the dust extinction at a lower level consistent with the map estimates does not change our conclusions, but would suggest a slightly lower mass for both the BH and its companion.}
 
When we relax the derived constraints on our fit -- i.e. suppressing constraints (iii) and (vi) -- we  find brighter systems than those preferred by our more conservative model selection. These would require more significant deviation from the expected, \Gaia-derived distance and line of sight extinction but permit systems that are at greater distances than those given here. They do not require a main sequence B-star and thus allow a very massive BH \citep[similar to the scenario found by][]{LB1}.

\subsection{The surface abundances of LB-1}.

\begin{table}
	\centering
	\caption{Summary of the best fitting abundances values using the strongly constrained values.}
	\label{tab:abundance}
	\begin{tabular}{cccc} 
	    \hline
	            & Initial Mass  & Best fitting  & \\
		Element & Fraction      & Mass fraction & $\log(X_f/X_i)$ \\
        \hline
        He & 0.280      &  0.795$\pm$0.009  &    0.45\\
        C & 3.46$\times 10^{-3}$     & (8.9$\pm$0.6)$\times 10^{-5}$ &   -1.59\\
        N & 1.05$\times 10^{-3}$     & 1.24$\pm$0.01$\times 10^{-2}$   &   1.07\\
        O & 9.64$\times 10^{-3}$     & 1.2$\pm$0.12$\times 10^{-3}$  &   -0.92\\
		\hline
	\end{tabular}
\end{table}

As mentioned above \citet{2020A&A...633L...5I} and \citet{2020A&A...634L...7S} re-analysed the optical spectrum of the companion star in LB-1. They also provided estimates of the surface abundances. Both show evidence of enhanced nitrogen and depleted carbon. \citet{2020A&A...633L...5I} however show that by mass fraction helium and nitrogen are enhanced by 0.4 dex and 0.7 dex respectively, while carbon and oxygen are depleted by -1.4 dex and -1 dex respectively. This is a similar pattern to that we derive from those models that match the other LB-1 parameters, which are shown in Table \ref{tab:abundance}. The agreement is another validation that the companion star is a post-binary interaction object and not a main sequence star. While the abundances are not a perfect match, the uncertainties in the measurement and the range of model abundances are significant. However if the observational abundance measurements could be refined then a more careful search through stellar evolution models of varying initial metallicity may lead to tighter evolutionary constraints on the LB-1 system.

\section{The Distance to LB-1}
\label{sec:distance}
The scenario for a 70\,$M_\odot$ BH demands that the source parallax reported by \Gaia\ is incorrect by a factor of two. \citet{LB1} argue that the astrometric wobble induced by the BH is sufficient to bias the parallax by this amount. This is only plausible if \Gaia\ preferentially observed the system at particular times during the system's orbit. Otherwise the wobble will induce variance to the astrometric fit, not bias. Taking the \Gaia\ scanning law\footnote{https://gaia.esac.esa.int/gost/} and a period of 78.9\, days \citep{LB1}, we find that \Gaia\ observed LB-1 at uniformly random times during the orbit. This makes the parallax unlikely to be systematically biased by the orbital motion.

\begin{figure*}
    \includegraphics[width=2\columnwidth]{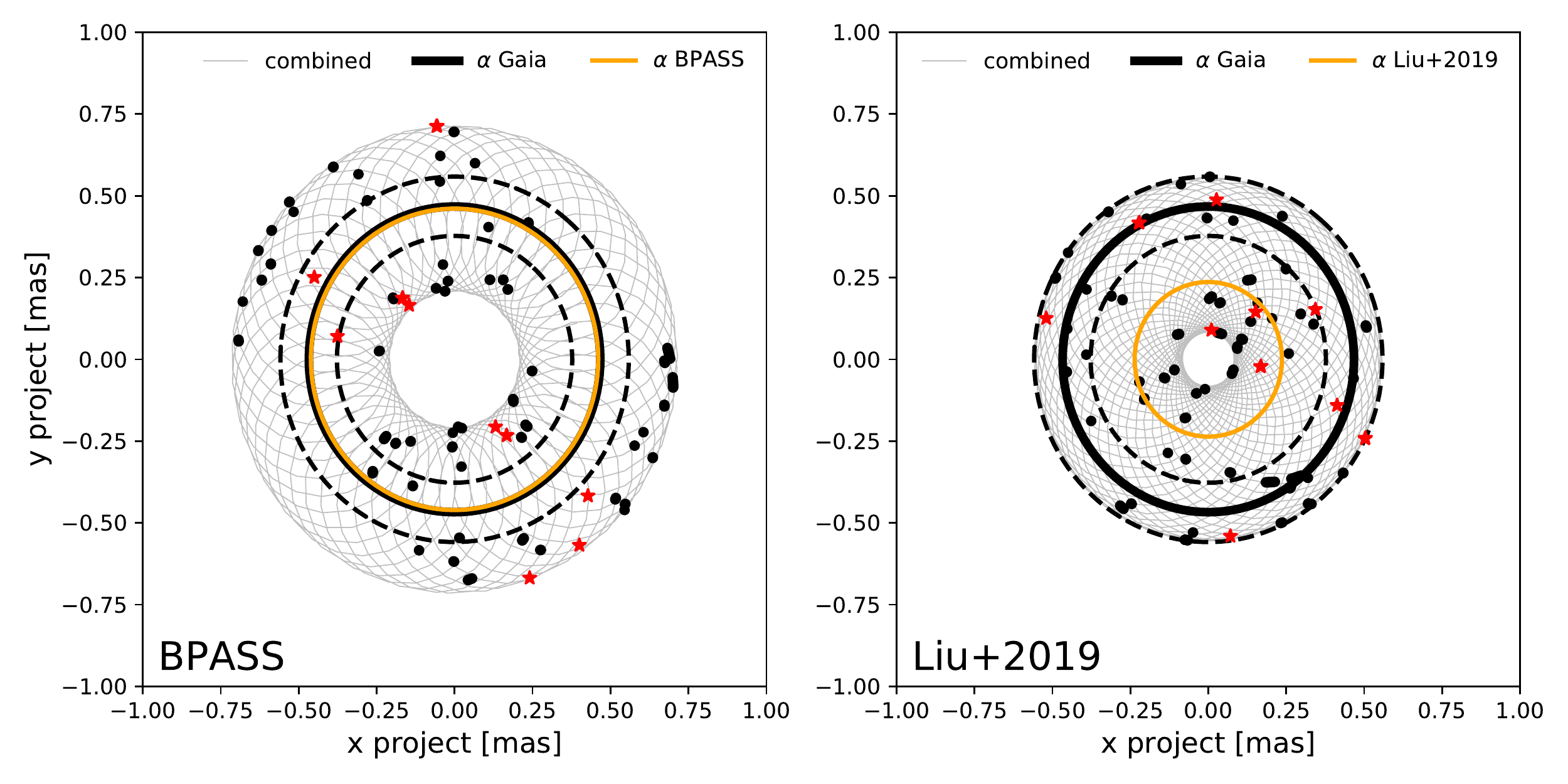}
    \caption{
    Projected size of the motion of LB-1 due to parallax based on derived distances (orange), Gaia's observed parallax and parallax error (black solid and dotted lines), and the combined effect (grey) of the parallax and binary motion for a randomly chosen orbital phase and face-on inclination based on the best fit results in Table 1 (left) and the derived results of Liu+2019 (right). Black dots show the position of the B-star at times of planned observations through 2025 based on the Gaia scanning pattern, while red stars show the observed B-star positions during the Gaia DR2 period.
    }
    \label{fig:orbits}
\end{figure*}

\begin{figure*}
    \includegraphics[width=2\columnwidth]{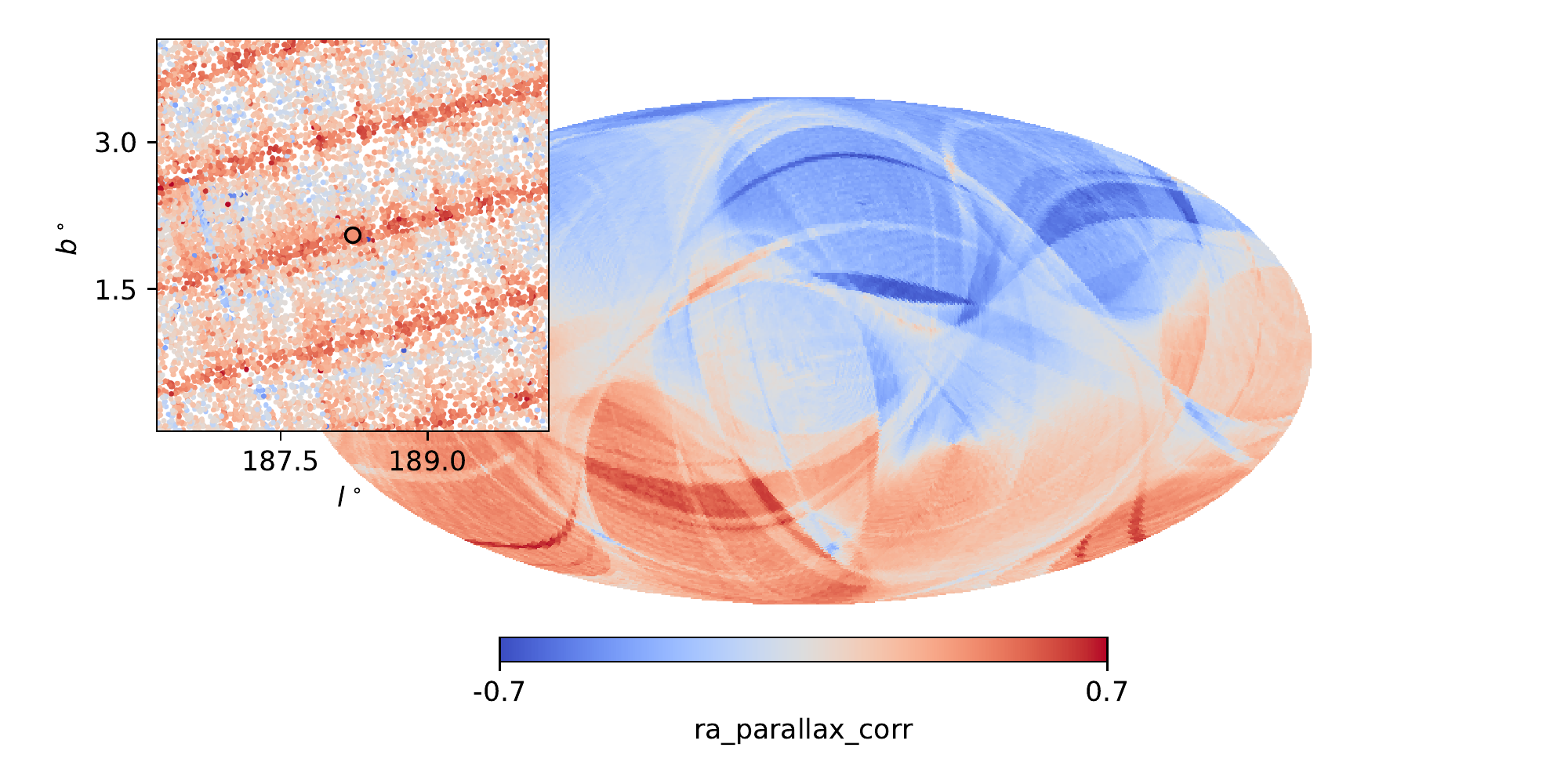}
    \caption{The mean coefficient of correlation (bounded between -1 and +1) in uncertainties of right ascension and parallax, as reported by Gaia, for all sources with G < 14. The correlation coefficient between right ascension and parallax for LB-1 (Gaia DR2) is 0.5397. The inset axes shows individual sources within $2^\circ$ of LB-1, where LB-1 is highlighted. The  \texttt{ra\_parallax\_corr} of LB-1 is entirely consistent with \Gaia's scanning pattern and not due to astrometric wobble.}
    \label{fig:ra_parallax_corr}
    \includegraphics[width=2\columnwidth]{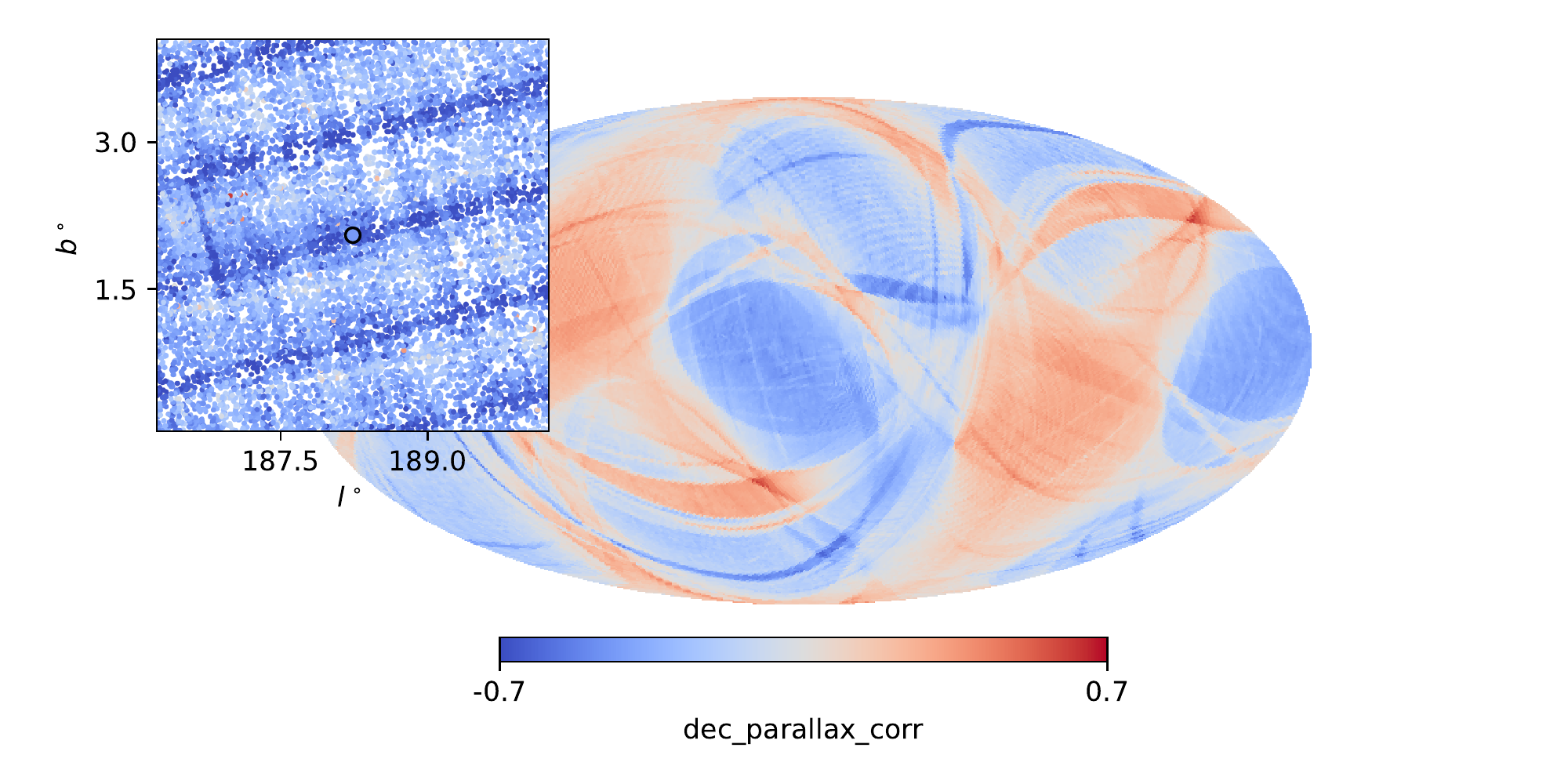}
    \caption{The mean correlation coefficient between declination and parallax for all \Gaia\ sources with $G < 14$. The correlation coefficient between declination and parallax for LB-1 (Gaia DR2) is -0.6187. Inset axes and colours are as per Figure~\ref{fig:ra_parallax_corr}. The \texttt{dec\_parallax\_corr} of LB-1 is entirely consistent with \Gaia's scanning pattern and not due to astrometric wobble.}
    \label{fig:dec_parallax_corr}
\end{figure*}

The projected semimajor axis of the B-star orbit based on the derived values of LB-1 by \citet{LB1} is $\sim0.322\,\rm{mas}$, while the parallax is $\sim0.236\,\rm{mas}$. This suggests that the position of the B-star observed by \Gaia\ should be significantly affected by the binary motion around the BH. However, based on the results in Table~\ref{tab:results}, at a distance of $2.17\,\rm{kpc}$ the projected semimajor axis of the B-star orbit is $\sim0.255\,\rm{mas}$ while the parallax is $\sim0.461\,\rm{mas}$. The parallax derived by \Gaia\ is $0.44\pm{0.086}\,\rm{mas}$ which suggests that the orbital motion should not have a significant effect on the observed parallax. This is inconsistent with the derived results of \citet{LB1}, but is not in strong tension with the results in Table~\ref{tab:results}. Figure~\ref{fig:orbits} illustrates the parallax from the inferred distance to LB-1 for our best fit model and that derived by \citet{LB1} as well as the parallax observed by \Gaia. An example of the combined parallax and binary motion, where a random orbital phase is  chosen, is also shown. A different phase choice will change the exact position of the companion star, but will not have a broad effect on the overall motion that the companion traces over long periods of time since LB-1's orbital period is significantly less than the observing span of the second \Gaia\ data release. The added binary motion of the companion star using our best fit model is consistent 
with the reported \Gaia\ parallax error. Conversely, the observed parallax is not well described by the combined orbital and parallactic motion of the companion star from \citet{LB1} because the combined motion is dominated by the orbit around the BH. We therefore find that a lower mass BH and companion star with a closer distance is favored by \Gaia's observed parallax over a higher mass BH and companion at a larger distance. We also note that while LB-1 was only visited $12$ times during \Gaia's $2^{\rm{nd}}$ data release, it is scheduled to be observed $\sim100$ times by the end of the \Gaia\ mission.

\citet{LB1} also argue that the parallax reported by \Gaia\ is unreliable because of strong correlations between position and parallax. However, these correlations depend primarily on \Gaia's scanning pattern and not the source orbital properties. This is illustrated in Figures \ref{fig:ra_parallax_corr} and \ref{fig:dec_parallax_corr} where we show the mean \texttt{ra\_parallax\_corr} and \texttt{dec\_parallax\_corr} for all \Gaia\ sources brighter than $G < 14$. The \texttt{ra\_parallax\_corr} (\texttt{dec\_parallax\_corr}) defines the correlation coefficient, bounded between -1 and +1, of the uncertainty in right ascension (declination) and the uncertainty in parallax \citep[e.g., see][]{Lindegren:2018}. In the figures, the scanning pattern is clearly evident. The inset axes shows individual \Gaia\ sources within 2$^\circ$ in $(l, b)$ of LB-1, the largest circle. While \citet{LB1} comment that the magnitudes of these correlation coefficients are large, they are fully consistent with what is expected from the scanning pattern. 

In short, the orbit is evenly sampled in time, the expected astrometric wobble in our preferred scenario is consistent with \Gaia\ observations, and the correlation coefficients between position and parallax are an expected consequence of the \Gaia\ scanning law. For these reasons we argue that the \Gaia\ parallax for LB-1 remains reliable, making the 70\,$M_\odot$ scenario untenable. For the companion to be the 8$M_{\odot}$ B star at this distance would require a significant amount of dust to obscure the star and reduce it to the observed magnitudes.

\section{Discussion \& Conclusions}
\label{sec:disc}

The binary systems containing black holes reported by \citet{LB1} and  \citet{2019Sci...366..637T} represent significant discoveries that will be key to our future understanding of how stars evolve, particularly those that are massive enough to form black holes in their death throes. These systems represent a previously unknown population of relatively quiet binary systems rather than the more easily-found class of interacting black hole binaries \citep[e.g][]{2010ApJ...725.1918O,2011ApJ...741..103F}.

We have searched the outputs of a binary population synthesis code to find binary evolution models that match the LB-1 black-hole binary observed by \citet{LB1}. The only models that were found to match the properties of the observed system contain a BH with significantly lower mass than reported in \citet{LB1} -- a scenario they discounted due to a suggested lack of such models in the stellar model grids they used.

It is possible that they were not able to identify such models due to their use of a rapid population synthesis code \citep[e.g][]{2008ApJS..174..223B,2019arXiv191112357B}. Although these methods are powerful, they lack the ability to follow the response of the star to mass loss that is possible in detailed binary evolution codes. This allowed us to find a match for the LB-1 system that was not a main-sequence star. In addition our detailed models calculate the composition throughout that star, enabling us to compare these (shown in Table \ref{tab:abundance}) to those measured by \citet{2020A&A...633L...5I}. They agree well showing enhanced helium and nitrogen abudance and depleted carbon and oxygen.

Furthermore many rapid population synthesis codes also use the \citet{2012ApJ...749...91F} rapid supernova mechanism prescription to estimate the remnant mass formed at core-collapse. By construction, this does not allow the formation of low mass black holes below $\sim5$M$_{\odot}$, and was suggested to agree with observed black hole binaries by \citet{2012ApJ...757...91B}. However, there is growing evidence that there might be a significant population black holes with masses below 5~M$_{\odot}$ \citep[e.g][]{2016MNRAS.458.3012W,LB1,2019Sci...366..637T,2019A&A...632A...3G} although the evidence is not conclusive yet and other explanations could increase the estimated masses \citep[e.g][]{2020A&A...636A..20W}. The situation will become clearer once the results of the 3rd LIGO/VIRGO observing run are published given the number of candidate mass-gap events that have been detected\footnote{https://gracedb.ligo.org/superevents/public/O3/}. If the mass-gap is filled with black holes it would suggest that the \citet{2012ApJ...749...91F} rapid supernova mechanism may need to be revisited, and that fewer assumptions should be made regarding the final fate of compact objects formed in the `Mass Gap'.

\begin{figure*}
	\includegraphics[width=2\columnwidth]{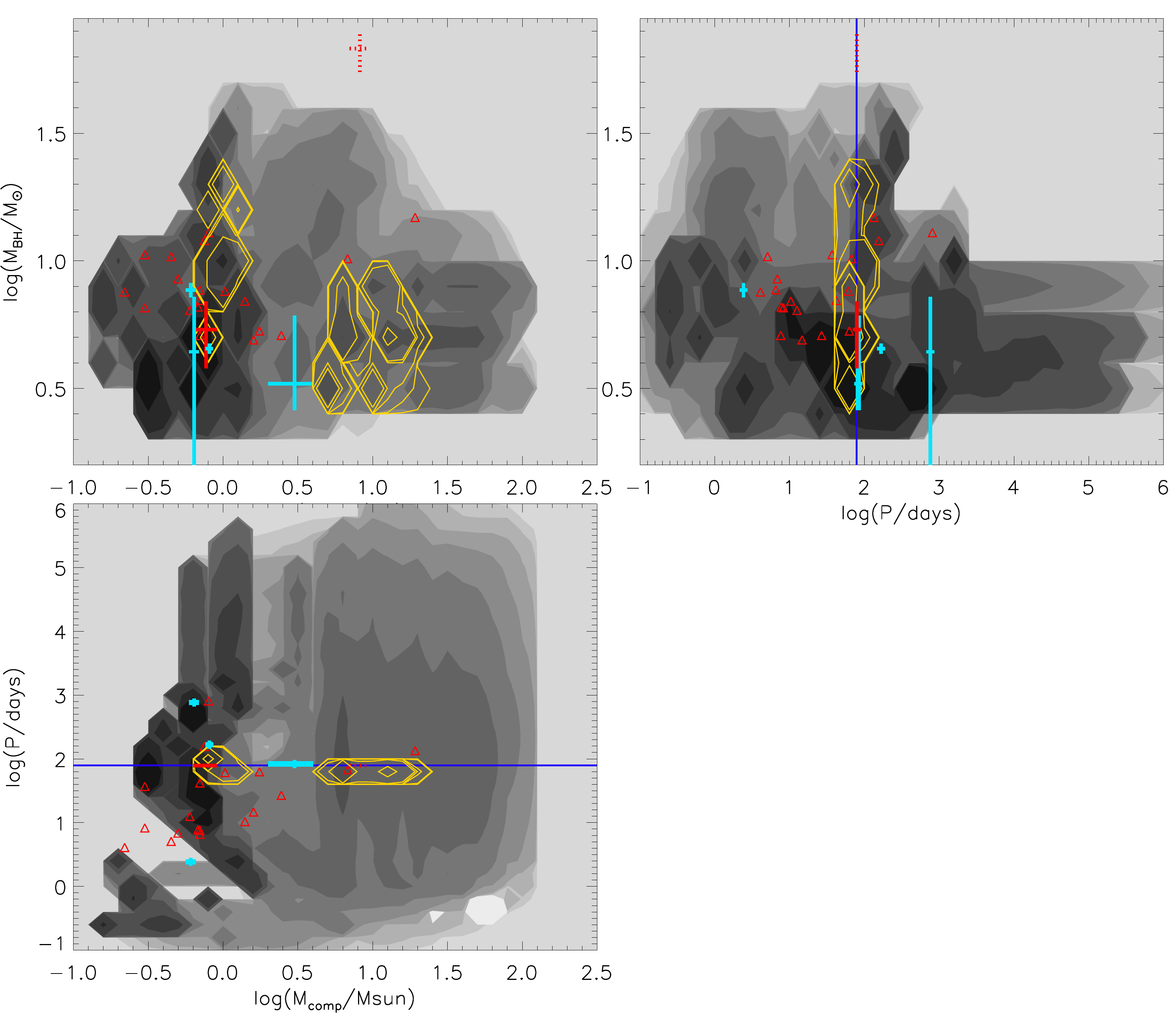}
    \caption{Prediction of compact remnant binaries at solar metallicity after 10 Gyrs of star formation. Each contour represents an order of magnitude in number of systems expected in that bin. The bin sizes are every 0.1 dex in mass and every 0.2 dex in $\log(P/{\rm days})$. The shaded contours represent all BH binary systems and the orange contours indicate those that match the observed companion in LB-1 and the orbital period. The dark blue line is the period of LB-1, the dotted red lines are the values for LB-1 determined by \citet{LB1}, the solid red lines are the values for LB-1 determined in this work and the light blue line are the values for the system observed by \citet{2019Sci...366..637T} and \citet{2019A&A...632A...3G}. The red triangles are values for the black-hole binaries listed by \url{https://stellarcollapse.org}. We note that if more metallicities were included the BPASS models would have models that match all observed systems. {Also high probability regions that have few observed systems are typically dominated by non-interacting black-hole binaries thus are less likely to have been observed to date compared to the X-ray selected systems plotted as red triangles.} }
    \label{fig:pop}
\end{figure*}

\citet{LB1} also reject lower mass B-stars due to their short evolutionary timescale. However, as we have discussed above, the number of these systems in our Galaxy while low, is not insignificant, making this discovery less unlikely than suggested. To further illustrate this point we show in Figure \ref{fig:pop} the full population of BHs with living companions expected from the BPASS population synthesis. As the figure illustrates, there are many similar periods to LB-1. 

Since systems such as LB-1 occur naturally within the BPASS fiducial population, we do not find its discovery surprising and do not need to invoke the existence of a 70\,M$_{\odot}$ BH that is in tension with our understanding of stellar evolution at the near-Solar metallicities expected within the Galaxy. We note that despite our reservations about the H$\alpha$ line tracing the BH dynamics, our favoured models would still imply a low value for $K_A$, consistent with the semi-amplitudes presented.

There are, of course, uncertainties that affect these predictions, primarily in how remnant masses are calculated from stellar models and the strength of any BH kicks. In addressing the former we note that the BPASS models are able to also reproduce the observed masses of BHs in X-ray binaries and also the masses and rates of GW transients \citep{2016MNRAS.462.3302E,2017PASA...34...58E,2019MNRAS.482..870E}. Additionally, BPASS already predicts larger BH masses at higher metallicities than other codes, as exemplified by the fact that it was possible in \citet{2016MNRAS.462.3302E} to reproduce the GW\,150914 system at a metallicity mass fraction of $Z=0.010$ (albeit at low probability). The maximum mass of a Solar metallicity BH at its formation in BPASS is 8\,M$_{\odot}$ but as we see in Figure \ref{fig:pop} this can grow up to $\approx$40\,M$_{\odot}$ as a result of accretion. Therefore achieving the proposed BH mass of 70\,M$_{\odot}$ is very difficult due to having to change details of mass transfer and core-collapse to try to make the black hole even more massive. We also note that trying to accrete more mass onto the black hole requires further super-Eddington accretion onto the black hole which may have observational consequences. This would also significantly affect the predicted mass distribution of BHs and gravitational wave transient rates from our BPASS models.

As we discuss above, LB-1 has a predicted orbital separation in our models of $\approx 200\,R_{\odot}$ (of the order of 1 AU). At a distance of 2.14\,kpc \citep{LB1} this would equate to an angular diameter of 0.255 milliarcseconds. This is certainly resolvable by \Gaia\ and future data releases will be able to firmly settle the nature of LB-1, including the BH mass \citep{Lindegren:2018, 2019ApJ...886...68A}. More such binaries will also be detectable in the future via \Gaia\ astrometry \citep[e.g.][]{2017MNRAS.470.2611M,2017ApJ...850L..13B,2019ApJ...878L...4B} and radial velocity surveys \citep[e.g.][]{2019ApJ...886...97Y}. While BH candidates discovered through radial velocity surveys will be subject to viewing angle uncertainties, astrometric discoveries by \Gaia\ will allow dynamical mass measurements down to the $5-10\%$ level \citep{2019ApJ...886...68A}.

{We note that during the review process of this article a recent careful analysis of the LB-1 system as be performed by \citet{2020arXiv200412882S} who find that LB-1 is very unlikely to contain a black hole at all. They find the the companion is in fact a Be star with the observed disk being that of the Be star that has accreted material from a companion. Despite this conclusion LB-1 is still an interesting system and future study is important to provide future firm constrains on the process of mass transfer in binary systems. For completeness of our study we have performed a search for models that fit the new parameters using constraints (i), (ii), (iv), (v), (vi) and the mass ratio, $M_2/M_1=4.7\pm 0.1$, \citep{2020arXiv200412882S} to select matching models within 3-$\sigma$ of these values. We show the resultant fit in Table \ref{tab:results2}. We see that the result is consistent with that in Table \ref{tab:results}, although the initial parameters for the system are now significantly different. For the current parameters they are similar but of course now with a star not a compact remnant. We suggest a more detailed study of this system in future once more is known of the nature of the Be star and with more detailed modelling of the how the secondary has responded to the mass transfer as our range of possible fits is wide.}

\begin{table}
	\centering
	\caption{Summary of the best fitting values assuming the parameters from \citet{2020arXiv200412882S}.}
	\label{tab:results2}
	\begin{tabular}{cccc} 
	    \hline
	    \multicolumn{2}{c}{Initial Parameters}&  \multicolumn{2}{c}{Current Parameters} \\
		\hline
        $M_{\rm i,1}/M_{\odot}$  &  4.51  $\pm$   0.62 & $M_{\rm 1}/M_{\odot}$  & 0.83 $\pm$ 0.16\\
        $M_{\rm i,2}/M_{\odot}$  &  2.98   $\pm$  0.66  & $M_{\rm 2}/M_{\odot}$& 3.79 $\pm$   0.80 \\
        $\log(P_i/{\rm days})$ &   1.05   $\pm$  0.13 &   $\log(P_{\rm fit}/{\rm days})$ &  1.88  $\pm$   0.05\\
      && $\log(L_1/L_{\odot})$ &  3.34 $\pm$   0.25 \\
         && $\log(L_2/L_{\odot})$ &  1.94 $\pm$   0.35 \\
           && $\log({\rm age/ yrs})$ &   8.18  $\pm$   0.14 \\
\hline
\end{tabular}
\end{table}

We emphasise that we have not varied any of our fiducial model parameters to find these binaries; they are naturally occurring in our population.
This same population has also been verified and tested against many other observations of individual stars, galaxies, galaxy populations and the thermal history of the Universe \citep[e.g][]{
2017PASA...34...58E,
2018PASA...35...49E,
2019MNRAS.482..870E,
2019PASA...36...41E,
2016MNRAS.459.3614M,
2018MNRAS.479..994R,
2016A&A...591A..22S,
2014MNRAS.444.3466S,
2016MNRAS.456..485S,
2018MNRAS.479...75S,
2016ApJ...826..159S,
2017MNRAS.469.2517W,2019MNRAS.490.5359W,
2016MNRAS.457.4296W,
2018MNRAS.477..904X,
2019MNRAS.482..384X}. 

In summary, our search of the currently available library of BPASS binary stellar evolution models has revealed a number of good matches to the observed LB-1 system.
We were able to identify at least one model whose stellar parameters place it at a distance in agreement with the \Gaia\ distance of $2.14^{+0.51}_{-0.35}$\,kpc. We find that the \Gaia\ distance should not be ignored due to the high values of the correlation coefficients. We estimate that several such systems are to be expected given the star formation history of the Milky Way.
We conclude that the LB-1 system of \citep{LB1} is likely to be explained by a naturally occurring binary containing a BH {\citep[or a normal star as suggested by][]{2020arXiv200412882S}} of moderate mass ($\approx 8M_{\odot}$), rather than one reaching 70$M_{\odot}$. If the higher mass is proven to be correct by further observations, we suggest that a re-evaluation of the accretion model with BPASS may be able to provide a more massive BH.

\section*{Acknowledgements}
The authors thank the referee for useful and careful comments that improved the paper. JJE and HFS acknowledge support from the University of Auckland and also the Royal Society Te Ap\={a}rangi of New Zealand under the Marsden Fund. ERS and DS acknowledge funding from the UK Science and Technology Research Council under grant ST/P000495/1. KB acknowledges support from the Jeffery L. Bishop Fellowship. ARC is supported in part by the Australian Research Council through a Discovery Early Career Researcher Award (DE190100656). Parts of this research were supported by the Australian Research Council Centre of Excellence for All Sky Astrophysics in 3 Dimensions (ASTRO 3D), through project number CE170100013.

We are grateful to Adrian Price-Whelan (Flatiron Institute) and all contributors to \texttt{The Joker}, \texttt{twobody}, \texttt{astropy}, and all their dependencies, for extensive documentation that helped facilitate this work. 

The BPASS project makes use of New Zealand eScience Infrastructure (NeSI) high performance computing facilities. New Zealand's national facilities are funded jointly by NeSI's collaborator institutions and through the Ministry of Business, Innovation \& Employment's Research Infrastructure programme. We also make use of the University of Warwick's Scientific Computing Research Technology Platform (SCRTP).

This work presents results from the European Space Agency (ESA) space mission Gaia. Gaia data is being processed by the Gaia Data Processing and Analysis Consortium (DPAC). Funding for the DPAC is provided by national institutions, in particular the institutions participating in the Gaia MultiLateral Agreement (MLA). The Gaia mission website is https://www.cosmos.esa.int/gaia . The Gaia archive website is https://archives.esac.esa.int/gaia .




\bibliographystyle{mnras}
\bibliography{refs} 








\bsp	
\label{lastpage}
\end{document}